\documentstyle[aps,pre,eqsecnum,multicol]{revtex}

\begin{document}

\title{Reaction-Controlled Diffusion}

\author{Steffen Trimper $^1$, Uwe C. T\"auber $^2$, 
                and Gunter M. Sch\"utz $^3$}
\address{$^1$ Fachbereich Physik, Martin--Luther--Universit\"at,
                D--06099 Halle, Germany \\
         $^2$ Department of Physics, Virginia Polytechnic Institute and State
                University, Blacksburg, VA 24061--0435 \\
         $^3$ Institut f\"ur Festk\"orperforschung, Forschungszentrum 
	      J\"ulich, D--52425 J\"ulich, Germany}
\draft
\date{\today}
\maketitle

\begin{abstract}
The dynamics of a coupled two-component nonequilibrium system is examined by
means of continuum field theory representing the corresponding master 
equation. Particles of species $A$ may perform hopping processes only when 
particles of different type $B$ are present in their environment. Species 
$B$ is subject to diffusion-limited reactions. If the density of $B$ 
particles attains a finite asymptotic value (active state), the $A$ species 
displays normal diffusion. On the other hand, if the $B$ density decays 
algebraically $\propto t^{-\alpha}$ at long times (inactive state), the 
effective attractive $A$-$B$ interaction is weakened. The combination of $B$ 
decay and activated $A$ hopping processes gives rise to anomalous diffusion, 
with mean-square displacement 
$\langle \vec x_A(t)^2 \rangle \propto t^{1 - \alpha}$ for $\alpha < 1$. 
Such algebraic subdiffusive behavior ensues for $n$-th order $B$
annihilation reactions ($n B \to \emptyset$) with $n \geq 3$, and $n = 2$ 
for $d < 2$. The mean-square displacement of the $A$ particles grows only 
logarithmically with time in the case of $B$ pair annihilation ($n = 2$) 
and $d \geq 2$ dimensions. For radioactive $B$ decay ($n = 1$), the $A$ 
particles remain localized. If the $A$ particles may hop spontaneously as 
well, or if additional random forces are present, the $A$-$B$ coupling 
becomes irrelevant, and conventional diffusion is recovered in the long-time 
limit.
\end{abstract}



\begin{multicols}{2}

\section{Introduction}

There has been considerable effort to elucidate the properties and conditions 
of anomalous diffusive behavior. A simple physical realization is given by 
diffusion on a fractal lattice \cite{fra}, where due to the increasing number 
of paths within the lattice, the time for a diffusion process will be 
prolongated. Also diffusion in random media with quenched disorder may be 
anomalous. Depending on the distribution of barrier heights (or depths of 
traps) one may observe normal diffusive or subdiffusive behavior, 
respectively, if the available number of diffusive paths is reduced by the 
presence of obstacles \cite{hav,hau}. Here we discuss a quite different 
situation in which diffusion is activated by the presence of particles or 
excitations which also propagate diffusively, but in the course of time 
decay. As a result the activated diffusion is rendered anomalous because the 
number of available paths decreases with time. However, the resulting 
structure of diffusive paths is not static, but evolves temporally. One may 
call this phenomenon dynamical fractality or dynamical disorder, depending 
on how the spatial distribution of excitations evolves in time.

We model this scenario by starting from a two-component system consisting of 
distinct particle species $A$ and $B$, with local time-dependent densities 
$\rho_A(\vec x, t)$ and $\rho_B(\vec x,t)$. An $A$ particle is allowed to 
perform hopping processes between adjacent neighboring sites on a lattice, 
provided there are one or more $B$ particles present in its vicinity. To be 
more specific, an $A$ particle hops from a site $j$ to a neighboring point 
$i$ subject to the condition that this site $i$ is already occupied with a 
particle of species $B$, and with a rate proportional to the local $B$ 
particle density. Obviously, such an effective attractive interaction
strongly influences the diffusive mobility of the $A$ species: Their 
mean-square displacement $\langle \vec x_A(t)^2 \rangle$ will depend on the 
time evolution of the local $B$ density $\rho_B(\vec x,t)$.

A non-trivial temporal behavior for $\rho_B(\vec x,t)$ will result if we 
submit the $B$ species locally to diffusion-limited reactions such as $n$-th
order annihilation $n B \to \emptyset$ (at the same or adjacent lattice 
points) or combined annihilation ($n \geq 2$) and spontaneous offspring 
production $B \to (m+1) B$ (the $B$ particles then perform branching and 
annihilating random walks, BARW). Once the time-dependence of $\rho_B(t)$ 
has been determined, we shall see that the $A$ kinetics is in the long-time 
limit to good approximation described on the basis of the associated 
mean-field rate equation. When the $B$ species is in an active state, i.e., 
$\rho_B(\vec x,t \to \infty) = \rho_B^\infty > 0$, with a basically 
homogeneous distribution in space, the $A$ particles will display normal 
diffusive behavior, with a diffusion constant $D_A \propto \rho_B^\infty$. 
In such a situation one has dynamical disorder, but there is always a finite 
fraction of sites available for hopping. However, an inactive phase, or the 
BARW critical point, are described either by an exponential decay 
$\rho_B(\vec x,t \to \infty) \propto e^{-\lambda t}$, in which case the $A$ 
particles remain localized, or by a power-law decrease 
$\rho_B(\vec x,t \to \infty) \propto t^{-\alpha}$ with a characteristic
exponent $\alpha > 0$. The diminishing density of $B$ particles reduces the 
induced mobility of the $A$ species, and these competing effects lead to 
subdiffusive behavior $\langle {\vec x}_A(t)^2 \rangle \propto t^{1 - \alpha}$ 
for $\alpha < 1$. In the borderline case $\alpha=1$ one has merely logarithmic 
growth $\langle {\vec x}_A(t)^2 \rangle \propto \ln t$.

Intuitively some of this behavior can be easily understood in the limit of
vanishing diffusion of the B-particles. We allow for multiple occupation, 
i.e., the site occupation number can be any integer between zero and infinity. 
When the occupation number of a $B$ particles at a certain lattice point is 
nonzero, that site is available for the $A$ species. In an inactive state the 
number of $B$ particles will be permanently reduced by reactions leading to a 
decreasing density of available sites for the $A$ species. This procedure can 
be viewed as an effective ``thinning out'' of lattice sites that leads to 
subdiffusive behavior for the $A$ particles reminiscent of (but distinct to) 
the mathematically considerably more complex phenomenon of diffusion on a 
fractal lattice.

\section{The model}

Here we present a more precise definition of our model in terms of a master 
equation which we formulate in the standard Fock-space formulation 
\cite{doi,gra,pel}, sometimes called ``quantum Hamiltonian formalism''
\cite{mat,sch}, particularly for particles with hard-core repulsion. 
Physically, this interaction is usually insignificant unless exclusion between 
different particle species or external driving forces need to be taken into 
account. This is intuitively clear for annihilating processes where the 
particle density tends to zero at late times \cite{pri,zho,sim,bel}, but
remains true also in the absence of particle reactions \cite{gsc} even in one
dimension. In some models, however, e.g., the annihilation-fission model 
\cite{how} or pair contact process with diffusion \cite{hen,hin}, site
occupation number constraints do play a crucial role. Hence, usually the 
specific choice of model is prescribed by the mathematical treatment used 
to analyze it. In the present context where we shall employ mean-field 
techniques and renormalization group arguments it is more advantageous to 
consider particles without site exclusion (for recent reviews, see e.g. 
Refs.~\cite{mat,car}).

We consider a system consisting of two different types of particles denoted as 
$A$ and $B$. The time evolution can be represented through an evolution 
operator $L$ \cite{doi,gra,pel}. The corresponding annihilation and creation 
operators are written as $a_i$ ($b_i$) and $a_i^\dagger$ ($b_i^\dagger$), 
where the index indicates a lattice point in $d$ space dimensions. For 
example, the normal hopping process of species $B$ from a site $j$ to its 
neighbor $i$ is described by the evolution operator 
$D \, (b_i^\dagger b_j - b_j^\dagger b_j)$, and for the entire lattice
therefore
\begin{equation}
  L_B = D \sum_{(ij)} \left( b_i^\dagger - b_j^\dagger \right)
        \left( b_j - b_i \right) \ ,
\label{evb}
\end{equation}
where $D$ is the hopping rate or diffusion constant.

An analogous expression would describe free diffusion of the $A$ particles. 
Here, however, we examine the situation that such a process is only allowed if 
there is at least one $B$ particle present at site $i$. If no representative 
of species $B$ is available at that site, an $A$ particle cannot move there. 
The time evolution operator for that process is proportional to
$(a_i^{\dagger} a_j - a_j^\dagger a_j) b_i^\dagger b_i$. The corresponding 
hopping process will occur provided an $A$ particle is in fact present at site 
$j$ and at least one $B$ particle occupies site $i$. Moreover, its rate is 
actually proportional to the number of $B$ particles present at site $i$. For 
the full system we obtain
\begin{equation}
  L_A = {\widetilde D} \sum_{(i,j)} \left( a_i^\dagger - a_j^\dagger \right)
   \left( a_j \, b_i^\dagger \, b_i - a_i \, b_j^\dagger \, b_j \right)  \ .
\label{eva}
\end{equation}
Here, ${\widetilde D}$ denotes the induced transition rate for the dynamical
process of species $A$.

In contrast to species $A$, the $B$ particles are subject to local reactions. 
A decreasing number of $B$ particles will lead to a slowing down for the 
motion of $A$'s through the lattice. For $n$-th order annihilation reactions 
$n B \to \emptyset$, the nonequilibrium evolution operator reads \cite{lee}
\begin{equation}
  L_R = \lambda_n \sum_i \left( 1 - {b_i^\dagger}^n \right) b_i^n \ .
\label{evr}
\end{equation}
Obviously, this operator describes the annihiliation of $n$ particles of type
$B$ at a lattice site $i$ provided such particles are available; $\lambda_n$
denotes the corresponding rate. Similarly, spontaneous branching processes 
$B \to (m + 1) B$ with rate $\sigma_m$ are described by \cite{cat}
\begin{equation}
  L_P = \sigma_m \sum_i \left( {b_i^\dagger}^m - 1 \right) b_i^\dagger b_i \ .
\label{evw}
\end{equation}
Together with Eq.~(\ref{evb}), $L_R$ and $L_P$ represent the time evolution
operator for branching annihilating random walks (BARW).

The complete dynamics is determined by $L= L_A + L_B + L_R (+ L_P)$, and may 
be encoded into a time-dependent ``state vector'' \cite{doi}
\begin{equation}
  | F(t) \rangle = \sum_{n_i} P(\vec n,t) \, | \vec n \rangle \ .
\label{ma2}
\end{equation}
Here $P(\vec n,t)$ is the evolving probability distribution for the 
unrestricted site occupation numbers $\vec n = \{ n_i \}$ for both $A$ and $B$
particles, and $| \vec n \rangle$ a basic vector containing all possible 
entries $n_i = 0,1,2,...\infty$, i.e., the eigenvalues of the second-quantized
bosonic particle number operators $a_i^\dagger a_i$ and $b_i^\dagger b_i$,
respectively. The state $| 0 \rangle$ represents the vacuum with no particles 
present, $a_i \, | 0 \rangle = 0 = b_i \, | 0 \rangle$. The state vector obeys 
the equation of motion
\begin{equation}
  \partial_t | F(t) \rangle = L \, | F(t) \rangle \ ,
\label{ma1}
\end{equation}
or formally $| F(t) \rangle = e^{L t} \, | F(0) \rangle$.

The nonequilibrium operator $L$ corresponds to, and is obtained from the
evolution operator $L'$ of the classical master equation that can generally be
written as
\begin{equation}
  \partial_t \, P(\vec n,t) = L' \, P(\vec n,t) \ ,
\label{ma3}
\end{equation}
and the matrix elements of $L$ and $L'$ are uniquely related to each other. 
The time-dependent average of an arbitrary physical quantity $G(\vec n)$ with 
the probability distribution $P(\vec n,t)$ can be cast into a ``matrix 
element'' form for the corresponding second-quantized operator $G(t)$ 
\begin{equation}
  \langle G(t) \rangle = \sum_{n_i} P(\vec n,t) \, G(\vec n) =
  \langle \Psi | \, G \, | F(t) \rangle \ ,
\label{ma4}
\end{equation}
with the projection state
$\langle \Psi | = \langle 0| \,\exp{\sum_i (a_i+b_i)}$.
Using the relation $\langle \Psi | \, L = 0$, the evolution equation for an
arbitrary operator $G$ becomes
\begin{equation}
  \partial_t \, \langle G \rangle = \langle \Psi |\, [G,L] \,| F(t) \rangle 
	\ .
\label{ma5}
\end{equation}
All the dynamical equations governing the classical problem are thus 
determined by the commutation rules of the underlying operators and the 
structure of the evolution operator $L$. In our case the dynamics of the model
is given by induced hopping processes for the $A$ particles and 
diffusion-limited reactions for the $B$ species, which we shall assume to be 
distributed randomly at the initial time $t = 0$. 

As a final step, we employ coherent basis states to represent the matrix 
element (\ref{ma4}) by means of a path integral \cite{pel,car}, and take the 
continuum limit. Absorbing factors containing the lattice constant into the 
diffusion and reaction rates, we may compute averages with a dynamical weight 
$\exp \bigl( - {\cal A}[\hat a,a,\hat b,b] \bigr)$ consisting of contributions
to the bosonic field action ${\cal A}$ describing the ordinary $B$ diffusion
\begin{equation}
  {\cal A}_B[\hat b,b] = \int \! d^dx \int \! dt \
        \hat b \left( \partial_t b - D \, \nabla^2 b \right) \ ,
\label{adb}
\end{equation}
the pure $n$-th order annihilation reactions
\begin{equation}
  {\cal A}_R[\hat b,b] = - \lambda_n \int \! d^dx \int \! dt
        \left( 1 - {\hat b}^n \right) b^n \ ,
\label{aaa}
\end{equation}
or offspring production processes,
\begin{equation}
  {\cal A}_P[\hat b,b] = \sigma_m \int \! d^dx \int \! dt
        \left( 1 - {\hat b}^m \right) \hat b \, b \ ,
\label{aaw}
\end{equation}
respectively. Finally the $A$ diffusion as induced by the coupling to the
$B$ species,
\begin{eqnarray}
  &&{\cal A}_A[\hat a,a,\hat b,b] = \! \int \!\! d^dx \! \int \!\! dt \, 
	\hat a \Bigl[ \partial_t a - {\widetilde D} \, (\nabla^2 a) \, 
	\hat b \, b \nonumber \\
  &&\qquad\qquad\qquad\qquad\qquad\qquad\quad\ + {\widetilde D} \, a \, 
	\nabla^2 (\hat b \, b) \Bigr] \ .
\label{ada}
\end{eqnarray}
Notice that $\hat b(\vec x,t) \, b(\vec x,t)$ represents the local density
$\rho_B(\vec x,t)$ (when appropriate ensemble averages are taken); the $A$ 
diffusion is thus mediated by the presence of $B$
particles. We remark that apart from the continuum limit, the mapping of the 
master equation onto the above field theory is exact and involves no further
approximations. (We have omitted the boundary contributions stemming from 
the initial conditions and the projection state here.)

\section{General considerations}

Clearly, the dynamic process for the $A$ particles as defined above is 
induced by the coupling to the reactive $B$ species only. When there are no 
$B$ particles present, $\rho_B(x,t) = 0$, the $A$ dynamics obviously ceases. 
Indeed, it turns out that there appears no noise in the dynamic equation 
governing the $A$ kinetics, which would formally appear as a contribution 
$\propto \hat a \, \hat a$ (or higher powers of $\hat a$) in the dynamic 
functional. In fact, any stochasticity emerges as a result of spatio-temporal 
fluctuations for the $B$ species (essentially reaction noise here). In order 
to further elucidate this point, we may derive effective Langevin-type 
equations for the local densities $\rho_A$ and $\rho_B$. To this end, we 
need to perform the shifts $\hat a = 1 + \widetilde a$, 
$\hat b = 1 + \widetilde b$, which take care of the annihilation operators 
appearing in the projection state $\langle \Psi |$, see Ref.~\cite{car}. To 
be specific, let us consider the case of $B$ pair annihilation reactions. 
Omitting temporal boundary terms describing the initial configuration, the 
new action becomes
\begin{eqnarray}
  &&{\cal A}[\hat a,a,\hat b,b] = \int \!\! d^dx \! \int \! dt \ \biggl[ 
	\widetilde a \Bigl( \partial_t a - D' \, \nabla^2 a - {\widetilde D} 
	\, (\nabla^2 a) \, b \nonumber \\
  &&\qquad\qquad + {\widetilde D} \, a \, (\nabla^2 b) \Bigr) - 
	{\widetilde D} \, \widetilde a \, (\nabla^2 a) \, \widetilde b \, b 
  	+ {\widetilde D} \, \widetilde a \, a \, \nabla^2 (\widetilde b \, b)
	\nonumber \\
  &&\qquad\qquad + \widetilde b \left( \partial_t b - D \nabla^2 b + 
	2 \lambda \, b^2 \right) + \lambda \, \widetilde b^2 b^2 \biggr] \ .
\label{sha}
\end{eqnarray}
Here, we have allowed for additional ordinary $A$ diffusion processes with 
rate $D'$. This dynamic action is equivalent to the following set of coupled 
Langevin equations,
\begin{eqnarray}
  &&\partial_t a = D' \, \nabla^2 a + {\widetilde D} \, (\nabla^2 a) \, b 
	- {\widetilde D} \, a \, (\nabla^2 b) + \zeta \ ,
\label{lda} \\
  &&\partial_t b = D \, \nabla^2 b - 2 \lambda \, b^2 + \eta \ ,
\label{ldb}
\end{eqnarray}
where the fluctuating forces with zero mean are characterized by the noise
correlations
\begin{eqnarray}
  &&\langle \zeta(\vec x,t) \, \zeta(\vec x',t') \rangle = 0 \ , \nonumber \\
  &&\langle \zeta(\vec x,t) \, \eta(\vec x',t') \rangle = {\widetilde D} \,
	[\nabla^2 a(\vec x,t)] \, b(\vec x,t) \, \delta(\vec x - \vec x') \,
	\delta(t-t') \nonumber \\
  &&\qquad\qquad\qquad\quad - {\widetilde D} \, a(\vec x,t) \, \nabla^2 
	[b(\vec x,t) \, \delta(\vec x - \vec x') \, \delta(t-t')] \ , 
	\nonumber \\
  &&\langle \eta(\vec x,t) \, \eta(\vec x',t') \rangle = - 2 \, \lambda \,
        b(\vec x,t)^2 \, \delta(\vec x - \vec x') \, \delta(t-t') \ .
\label{noi}
\end{eqnarray}
Taking averages, we may then identify $\rho_A(t) = \langle a(\vec x,t) 
\rangle$ and $\rho_B(t) = \langle b(\vec x,t) \rangle$, as Eqs.~(\ref{lda}) 
and (\ref{ldb}) obviously generalize the mean-field rate equations for the 
local particle densities. The reaction noise for the $B$ species displays the 
characteristic negative correlations (``imaginary noise''), which reflect the 
particle {\em anti}correlations induced by the annihilation reaction 
\cite{lee,adr,car,how}. When there are no $B$ particles left 
[$b(\vec x,t) = 0$], the fluctuations cease, characteristic of an absorbing 
inactive state. As anticipated, no noise contributions exist for the pure $A$ 
dynamics, but there appear $A$-$B$ noise cross-correlations. (Notice that pure 
diffusion noise does not appear explicitly here.)

Next, let us study what happens when the $A$ particles are subject to an
additional random force that leads to ordinary diffusion, i.e., the term
$\propto D'$ in the action (\ref{sha}). Obviously, one should expect that the 
induced diffusion $\propto \widetilde D$ is suppressed in this situation, and 
in the long-time limit standard diffusion prevails. This becomes indeed clear 
through simple power counting, introducing a momentum scale $\kappa$, i.e., 
$[x] = \kappa^{-1}$, and measuring time scales as $[t] = \kappa^{-2}$, as 
appropriate for diffusive dynamics. Then $[D] = [D'] = \kappa^0$ become 
dimensionless, and we infer the field scaling dimensions 
$[\hat a] = [\hat b] = [\widetilde a] = [\widetilde b] = \kappa^0$ and 
$[a] = [b] = \kappa^d$, as to be expected for $d$-dimensional particle 
densities. The remaining couplings (reaction rates) acquire the scaling 
dimensions
\begin{equation}
  [\sigma_m] = \kappa^2 \ , \ [\lambda_n] = \kappa^{2 - (n-1) \, d} \ , \
        [\widetilde D] = \kappa^{-d} \ .
\label{scd}
\end{equation}

A positive scaling dimension means that the corresponding parameter is 
relevant in the renormalization-group (RG) sense. E.g., the branching rate 
$\sigma_m$ carries the dimensions of a ``mass'' term, and indeed represents 
the decisive control parameter for BARW: In mean-field theory, the critical 
point must be at $\sigma_m = 0$, and is therefore described by the pure 
annihilation model, while for any positive $\sigma_m$ there will be only an 
active phase characterized by exponential correlations. The annihilation rate 
is relevant for $d < 2 / (n-1)$ dimensions, and irrelevant for $d > 2 / (n-1)$ 
\cite{cat}. Hence we identify the upper critical dimension, below which 
fluctuations in fact dominate the asymptotic behavior, as $d_c(n) = 2 / (n-1)$
for $n$-th order annihilation processes \cite{kra,lee}. Thus, for $n > 3$ 
fluctuations are not too important in any physical dimension $d \geq 1$.

Furthermore we notice that the coupling $\widetilde D$ is {\em irrelevant}, 
i.e., compared to the other parameters in the theory its influence should 
become negligible in the asymptotic long-time, long-wavelength limit. 
Evidently, $\rho_B(\vec x,t)$ either vanishes (inactive phase) or approaches a 
constant $\rho_B^\infty$ (active phase) as $t \to \infty$. In the former case, 
normal $B$ diffusion, if present ($D' > 0$), will dominate; in the latter 
situation, the combined quantity $\widetilde D \, \rho_B^\infty$ will 
effectively act as an ordinary diffusion constant, numerically renormalizing 
$D'$. In any case, we see that the ordinary $A$ diffusion process is not 
qualitatively affected by the induced hopping through attractive coupling to 
the $B$ density, and the associated noise cross-correlations. Also when 
$D' = 0$, as in our original model, and in a system with an initially 
{\em finite} number of $B$ particles, {\em asymptotically} the $A$ particles 
either remain localized or display standard diffusion. In this respect, in 
numerical simulations the induced anomalous diffusion in which we are 
interested here would appear as a {\em crossover} feature in the long-time 
kinetics and correspond to corrections to scaling to the leading asymptotic 
time dependence. In an infinite system, however, with initially finite $B$
{\em density}, the anomalous diffusion regime will persist indefinitely.

A corollary of these observations is that the rate $\widetilde D$ does not
acquire any non-trivial frequency or time dependence in the infrared. In the 
field theory language, we note that neither diffusive propagator for the
$A$ or $B$ species can be renormalized by the $(\hat a \hat b b a)$ four-point
vertex in the unshifted action (\ref{ada}), or equivalently, the three- and
four-point vertices in the shifted action (\ref{sha}). Consequently, the 
renormalization for the vertex functions $\Gamma_{\hat a \hat b b a}$ or 
$\Gamma_{\tilde a \tilde b b a}$ and $\Gamma_{\tilde a b a}$, respectively, 
can be determined to {\em all} orders in the perturbation expansion (with 
respect to $\widetilde D$) by means of a Bethe--Salpeter equation, or 
equivalently, a geometric series of loops containing just the $A$ and $B$ 
propagator. This leads to the renormalized wavevector- and frequency-dependent 
coupling
\begin{equation}
  \widetilde D_R(\vec q, \omega) = \widetilde D \Biggl[ q^2 + {\widetilde D} 
	\int \!\! \frac{d^dp}{(2 \pi)^d} \ \frac{p^2 \, [(\vec q - \vec p)^2 
	- p^2]}{-i \omega + D' p^2 + D p^2} \Biggr]^{-1} \ ,
\label{red}
\end{equation}
where ${\vec q}$ and $\omega$ denote the momentum and frequency transfer 
between the $A$ and $B$ particles. We may now set $D' = 0$ again, and 
investigate the long-wavelength limit $\vec q \to 0$,
\begin{equation}
  \frac{\partial}{\partial q^2} \widetilde D_R(\vec q, \omega)
	\bigg\vert_{q=0} \!\!\! = \widetilde D \left[ 1 + 
	\frac{\widetilde D}{D} \left( \frac{\omega}{D} \right)^{\!\frac{d}{2}} 
	\!\! \int \!\! \frac{d^dk}{(2 \pi)^d} \frac{k^2}{-i + k^2} 
	\right]^{-1} ,
\label{scl}
\end{equation}
where $k^2 = D p^2 / \omega$. Thus, as $\omega \to 0$, the fluctuation 
corrections vanish (provided the integral is regularized in the ultraviolet 
with an appropriate cutoff), and the renormalized coefficient $\widetilde D_R$ 
in Eq.~(\ref{scl}) approaches the original ``bare'' constant $\widetilde D$.

This is to be contrasted with the infrared-singular behavior of e.g. the $B$
pair annihilation rate, for which an analogous procedure yields \cite{lee}
\begin{eqnarray}
  &&\lambda_R(\vec q, \omega) = \lambda \Biggl[ 1 + \frac{\lambda}{D} \! \int
        \!\! \frac{d^dp}{(2 \pi)^d} \ \frac{1}{-i \omega / D + q^2 / 4 + p^2}
        \Biggr]^{-1} \ , \nonumber \\
  &&\lambda_R(0, \omega) = \lambda \Biggl[ 1 + \frac{\lambda}{D} \left(
        \frac{\omega}{D} \right)^{\!\frac{d-2}{2}} \!\! \int \!\!
 \frac{d^dk}{(2 \pi)^d} \, \frac{1}{-i + k^2} \Biggr]^{-1} .
\label{anr}
\end{eqnarray}
For $d > d_c(2) = 2$, again $\lambda_R(0,0) = \lambda$ is just the original
rate constant, resulting in the mean-field power law
$\rho_B(t) \propto t^{-1}$. However, for $d < d_c(2) = 2$, 
$\lambda_R(0,\omega) \propto \omega^{1-d/2}$ vanishes for low frequencies.
Inserting the corresponding effective time-dependent rate 
$\lambda_R(t) \propto t^{-1+d/2}$ into Eq.~(\ref{ldb}) leads to the correct
slower algebraic decay $\rho_B(t) \propto t^{-d/2}$.

In summary, the $B$ process itself is, per definition of our model, not 
influenced by the $A$ dynamics. In the renormalization group treatment, this 
is reflected by the fact that the coupling $\widetilde D$ is irrelevant, and 
thus does not affect the long-time behavior. Yet the induced hopping rate 
$\widetilde D$ is of course crucial for the $A$ species kinetics, and must 
be kept even in the mean-field approximation. We may thus solve for the $B$ 
kinetics first, and then explore its influence on the induced $A$ diffusion. 
Henceforth, we shall again set $D' = 0$, as otherwise simple ordinary $A$ 
diffusion would ensue, with $\widetilde D$ then irrelevant also for the $A$ 
kinetics, and the entire coupling of the $A$ and $B$ processes would 
disappear asymptotically. In the following, we shall study the $A$ kinetics 
assuming a spatially homogeneous, but time-dependent distribution of $B$ 
particles, which leads us to a mean-field description.

\section{Mean-field evolution equations}

\subsection{Annihilation kinetics}

Let us assume we can ignore spatial fluctuations for the $B$ species entirely,
and ignore the reaction noise. For the $n$-th order annihilation processes, we 
saw that this as at least a qualitatively correct description for 
$d > d_c(n) = 2 / (n-1)$, i.e., for $d > 2$ in the case of pair annihilations, 
$d > 1$ for the third-order process $3 B \to \emptyset$, and in any physical 
dimension for $n > 3$. The evolution equation can either be obtained directly 
from the non-equilibrium operator $L_R$ in Eq.~(\ref{evr}) and the equation of 
motion (\ref{ma5}), or from solving for the stationarity condition 
$\delta {\cal A}/\delta \hat b = 0$ for the action 
${\cal A} = {\cal A}_B + {\cal A}_R$, Eqs.~(\ref{adb}) and (\ref{aaa}), 
setting $D = 0$. (Notice that $\delta {\cal A}/\delta b = 0$ is always solved 
by $\hat b = 1$.) Either procedure results in the obvious mean-field rate 
equation
\begin{equation}
  \partial_t \, \rho_B(t) = - n \lambda_n \, \rho_B(t)^n \ ,
\label{mfb}
\end{equation}
which is readily integrated for $n > 1$,
\begin{equation}
  \rho_B(t) = \frac{\rho_B(0)}{\left( 1 + t / \tau \right)^{1/(n-1)}} \ , 
	\quad \tau = \frac{\rho_B(0)^{1-n}}{n(n-1) \, \lambda_n} \ ,
\label{mas}
\end{equation}
i.e., for $t \gg \tau$ the $B$ density decays algebraically 
$\propto t^{-1/(n-1)}$ in this approximation, while of course for $n = 1$
\begin{equation}
  \rho_B(t) = \rho_B(0) \, e^{- \lambda_1 t} \ .
\label{mrd}
\end{equation}

In the same manner, we may obtain the evolution equation for the $A$ species,
or just consider Eq.~(\ref{lda}) for $D' = 0$ and vanishing noise. In the 
spirit of mean-field theory, we assume a homogeneous $B$ density, and obtain
\begin{equation}
  \partial_t \, \rho_A(\vec x, t) = \widetilde D \, \rho_B(t) \, \nabla^2
        \rho_A(\vec x, t) \ .
\label{mfa}
\end{equation}
Again, this equation can be solved exactly, considering a delta-like density
distribution for the $A$ species at the initial time $t=0$. As in this 
mean-field approach the $\rho_B(t)$ is assumed to be spatially uniform, the 
$A$ species will be Gaussian distributed in space, just like in ordinary 
diffusion,
\begin{equation}
  \rho_A(\vec x, t) = \left( \frac{1}{2 \pi \, \langle \vec x_A^2(t) \rangle}
        \right)^{d/2} \exp \left(- \frac{\vec x^2}{2 \, \langle \vec x_A^2(t)
        \rangle} \right) \ .
\label{gau}
\end{equation}
However, the $B$ decay (or lattice depletion) will be reflected in the 
anomalous time-dependence of the width (mean-square displacement). A 
straightforward brief calculation yields
\begin{equation}
  \left\langle \vec x_A^2(t) \right\rangle = 2 \widetilde D \int_0^t \!
        \rho_B(t') \, dt' \ .
\label{dis}
\end{equation}

For $n=1$, i.e., the simple exponential decay (\ref{mrd}), the result is
\begin{equation}
  \left\langle \vec x_A^2(t) \right\rangle =
        \frac{2 \widetilde D \, \rho_B(0)}{\lambda_1} \left( 1 -
        e^{-\lambda_1 t} \right) \ .
\label{xrd}
\end{equation}
Initially ($\lambda_1 t \ll 1$) one finds normal diffusion with effective
diffusion constant $\bar D = \widetilde D \, \rho_B(0)$, but at long times the 
mean-square displacement approaches a constant, and the $A$ particles remain 
localized in a region of volume 
$\propto \langle \vec x_A^2(t \to \infty) \rangle^{d/2} = (2 \widetilde D \,
\rho_B(0) / \lambda_1)^{d/2}$. Given that this simple process is characterized 
by short-range correlations in space and time only, we do not expect any 
considerable modification through fluctuation effects.

In the pair annihilation case, $n=2$, one finds
\begin{equation}
  \left\langle \vec x_A^2(t) \right\rangle = 2 \widetilde D \, \rho_B(0) \
        \ln \left( 1 + \frac{t}{\tau} \right) \ ,
\label{xpa}
\end{equation}
while the mean-field result for $n > 2$ reads
\begin{equation}
  \left\langle \vec x_A^2(t) \right\rangle = 2 \widetilde D \, \rho_B(0) \,
	\frac{n-1}{n-2} \, \tau \left[ \left( 1 + \frac{t}{\tau}
        \right)^{\frac{n-2}{n-1}} - 1 \right] \ .
\label{xna}
\end{equation}
In the asymptotic regime $t \gg \tau$, this implies anomalous diffusion
according to
\begin{equation}
  \left\langle \vec x_A^2(t) \right\rangle \propto t^{2 / (2 + \Theta)}
\label{and}
\end{equation}
with a positive exponent $\Theta = 2 / (n-2)$ indicating {\em subdiffusive}
behavior. In the limit $n \to \infty$ we have $\Theta \to 0$, and conventional 
diffusion is recovered. The reason is of course that for large $n$ the 
depleting reactions become very unlikely, as $n$ particles are required to 
meet at the same lattice site. Thus, low-order $B$ species reactions are much 
more effective in slowing down the $A$ diffusion. The time scale for the 
crossover to the pure algebraic decay of the $B$ particle density and 
subsequently for the anomalous $A$ diffusion is given by 
$\tau \propto \rho_B(0)^{1-n} / \lambda_n$. The crossover to the asymptotic 
slow dynamics is fast for large initial densities and reaction rates.

The above analysis should be qualitatively correct for $n > 3$, as the
corresponding critical dimension $d_c(n) < 1$. For $n=2$, i.e., $B$ pair 
annihilation processes in $d \leq 2$ dimensions, we know that at long times 
{\em anti}correlations develop \cite{adr,lee,car}: Initially close-by 
particles disappear quickly, and only widely separated ones survive. This 
effective ``repulsion'' should result in a roughly uniform spatial $B$ 
distribution even for a clustered initial configuration. Given that the 
coupling coefficient $\widetilde D$ itself does not renormalize, we therefore 
expect that our decoupling assumption leading to Eq.~(\ref{mfa}) should 
represent a fair approximation, provided the correct time dependence of the 
$B$ density is inserted. For $d < 2$, the asymptotic result is
\begin{equation}
  \rho_B(t) \propto t^{-d/2} \ ,
\label{pad}
\end{equation}
see Ref.~\cite{lee} and also Sec.~III following Eq.~(\ref{anr}), whence
\begin{equation}
  \left\langle \vec x_A^2(t) \right\rangle = 2 \bar D \, t^{1 - d / 2}
\label{xda}
\end{equation}
with an appropriate effective rate $\bar D \propto \widetilde D / (1 - d/2)$.
In low dimensions, this algebraic subdiffusive behavior with
$\Theta = 2d / (2-d)$ replaces the logarithmic law (\ref{xpa}). At the 
critical dimensions $d_c(2) = 2$, one finds the typical logarithmic 
corrections 
\cite{lee}
\begin{equation}
   \rho_B(t) \propto t^{-1} \, \ln t \ ,
\label{pa2}
\end{equation}
implying
\begin{equation}
  \left\langle \vec x_A^2(t) \right\rangle \propto {\widetilde D} \, 
  	(\ln t)^2 \ ,
\label{xa2}
\end{equation}
which also describes slower kinetics than given by the mean-field result
(\ref{xpa}). For the case of $n = 3$ at its critical dimension $d_c(3) = 1$,
\begin{equation}
  \rho_B(t) \propto (t^{-1} \, \ln t)^{1/2} \ , 
\label{ta1}
\end{equation}
and one would therefore expect the leading time dependence
\begin{equation}
  \left\langle \vec x_A^2(t) \right\rangle \propto {\widetilde D} \, 
  	(t \, \ln t)^{1/2} \ ,
\label{xa3} 
\end{equation}
i.e., essentially a square-root power law with logarithmic corrections.

\subsection{BARW kinetics}

We now extend the $B$ dynamics and include branching processes of the form
$B \to (m+1) \, B$ with rate $\sigma_m$, described by Eqs.~(\ref{evw}) or
(\ref{aaw}). The mean-field rate equation (\ref{mfb}), with $n \geq 2$, is 
then replaced by 
\begin{equation}
  \partial_t \rho_B(t) = - n \lambda_n \, \rho_B(t)^n + m \sigma_m \, 
	\rho_B(t) \ ,
\label{bra1}
\end{equation}
which has two stationary solutions $\rho_B = 0$ (inactive phase) and
\begin{equation}
  \rho_B^\infty = \left( \frac{m \sigma_m}{n \lambda_n} \right)^{1/(n-1)}
\label{rbs}
\end{equation}
(active phase). For any $\sigma_m > 0$, the latter turns out to be stable, 
i.e., BARW are always in the active phase in the mean-field approximation. The 
explicit solution of Eq.~(\ref{rbs}) furthermore shows that the asymptotic 
density $\rho_B^\infty$ is exponentially approached,
\begin{equation}
  \rho_B(t) = \frac{\rho_B^\infty}{\left[ 1 + C \, e^{- (n-1) m \sigma_m t}
  \right]^{1/(n-1)}} \ ,
\label{bam}
\end{equation}
where $C = [\rho_B^\infty / \rho_B(0)]^{n-1} - 1$. Again, Eq.~(\ref{mfa}) is 
solved by the Gaussian distribution (\ref{gau}) with mean-square displacement 
(\ref{dis}). The ensuing integral is readily calculated for some special 
cases, e.g., for $n = 2$
\begin{equation}
  \left\langle \vec x_A^2(t) \right\rangle = 2 \widetilde D \, \rho_B^\infty
        \left[ t + \frac{1}{m \sigma_m} \ln\!\left( 1 + C \, 
	e^{- m \sigma_m t} \right) \right] \ ,
\label{ba2}
\end{equation}
whereas for $n=3$
\begin{equation}
  \left\langle \vec x_A^2(t) \right\rangle = \frac{\widetilde D \rho_B^\infty}
        {m \sigma_m} \ \ln \!\left( \frac{\sqrt{1 + C \, e^{- 2 m \sigma_m t}}
        + 1} {\sqrt{1 + C \, e^{- 2 m \sigma_m t}} - 1} \right) \ .
\label{ba3}
\end{equation}
In general, asymptotically normal diffusion with effective diffusion 
coefficient $\widetilde D \, \rho_B^\infty$ is recovered in the active state,
\begin{equation}
  \left\langle \vec x_A^2(t) \right\rangle = 2 \widetilde D \, \rho_B^\infty 
	\, t \ .
\label{mbw}
\end{equation}
The properties of the active phase with an asymptotically homogeneous $B$
density are not much influenced by fluctuations, and hence Eq.~(\ref{mbw})
should aptly describe the ensuing $A$ kinetics even beyond mean-field theory.

For the possible existence of an inactive phase, and the characterization of
the ensuing critical behavior, fluctuation effects are however of utmost 
importance for $n = 2$, and it turns out that the cases of odd and even
offspring number $m$ need to be distinguished. For odd $m$, aside from all 
lower-order branchings, first-order decay processes $B \to \emptyset$ are 
generated, and become sufficiently efficient to shift the critical point to 
$\sigma_c > 0$ for $d \leq 2$ dimensions. The emerging transition at 
$\sigma_c > 0$ can be shown to be in the generic directed-percolation (DP) 
universality class \cite{cat}. The inactive phase is then governed by 
exponential $B$ density decay, whereupon the $A$ species will become localized 
according to Eq.~(\ref{xrd}). At the critical point itself, the $B$ species 
density decays according to a power law $\rho_B(\vec x,t) \sim t^{-\alpha}$, 
with $\alpha = \beta / z \nu_\perp$ given by DP critical exponents in $d=1$ 
and $d=2$, respectively. This would suggest 
$\langle \vec x_A^2(t) \rangle \propto t^{1 - \alpha}$; yet the $B$ density 
is far from uniform at the critical point, and is instead characterized by 
the appearance of {\em fractal} density clusters. While we would still expect 
subdiffusive behavior for the $A$ species with $\Theta > 0$, this exponent 
will likely be influenced by the power-law correlations in the critical $B$ 
density. In the case of even $m$, for which the $B$ particle number parity is 
locally conserved under the reactions, a non-trivial transition with 
$\sigma_c > 0$ is possible only for $d \leq d_c' \approx 4/3$ dimensions. The 
inactive phase is then given by the pure pair annihilation theory, and 
consequently Eq.~(\ref{xda}) should provide a fair description for the ensuing 
anomalous $A$ diffusion. The critical behavior is governed by a different 
parity-conserving universality class, with $\alpha < 1/2$. In this instance, 
we again expect the above mean-field description to be rather inaccurate.

\section{Conclusions}

We have studied a novel mechanism to induce anomalous diffusion. Whenever an 
active particle of the $A$ species performs a random walk on a lattice, it may 
visit a certain lattice site only provided this site is already occupied by at 
least one $B$ particle. The random walk is prolongated when the $B$ particles 
react with each other in such a manner that the $B$ species density is 
decreasing. If that decay is exponential (first-order reaction), then after a 
short time interval (given by the inverse decay rate) the $B$ species has 
disappeared and a further visit of an $A$ particle at that site is impossible.
As a consequence the $A$ species, after some initial mixing, remains 
localized. When the $B$ species undergo reactions of higher order, requiring 
at least two $B$ particles to meet at a lattice site, an algebraic decay 
ensues that allows hopping processes for the $A$ species to occur for a much 
longer period. However, the random walk process is slowed down considerably as 
the $B$ density diminishes, resulting in a much shorter mean-square 
displacement of $A$ particles as compared with conventional diffusion. The 
emerging anomalous diffusion is governed by power laws or logarithmic behavior 
that can (approximately) be related to the asymptotic time behavior of the 
reacting $B$ particle density. In this instance one may view this process as 
resembling diffusion on a dynamical fractal. Only when at long times the $B$ 
density remains finite and nearly homogeneous, conventional $A$ diffusion is 
recovered. This situation corresponds to diffusion with dynamical disorder,
where in the long-time limit the $B$ particles, with largely decayed 
fluctuations, merely resemble a quasi-static inhomogeneous background for the
$A$ kinetics. The consistent mathematical treatment of diffusion on a static 
fractal, as well as induced diffusion processes on critical (isotropic or 
directed) percolation clusters or near BARW critical points remains an open 
problem that requires more sophisticated analysis beyond the largely 
mean-field approach presented here.

\acknowledgements
We acknowledge fruitful discussions with Hans-Karl Janssen, Beate Schmittmann,
and Olaf Stenull.

\end{multicols}
\end{document}